\documentclass[aip,jcp,amsmath,amssymb,reprint]{revtex4-2}

\usepackage{graphicx}
\usepackage{dcolumn}
\usepackage{bm}
\usepackage[utf8]{inputenc}
\usepackage[T1]{fontenc}
\usepackage{mathptmx}
\usepackage{natbib}
\usepackage{enumitem}   
\usepackage[table,xcdraw]{xcolor}
\usepackage{array}
\usepackage{amsmath, lipsum}
\usepackage{physics}

\usepackage[version=3]{mhchem}
\usepackage{etoolbox}
\usepackage{booktabs}
\usepackage{pgfplots}
\pgfplotsset{compat=newest}

\definecolor{color1}{HTML}{468B97}
\definecolor{color2}{HTML}{8EAC50}
\definecolor{color3}{HTML}{EF6262}

\definecolor{crimson2143940}{RGB}{214,39,40}
\definecolor{darkgray176}{RGB}{176,176,176}
\definecolor{darkorange25512714}{RGB}{255,127,14}
\definecolor{darkturquoise23190207}{RGB}{23,190,207}
\definecolor{forestgreen4416044}{RGB}{44,160,44}
\definecolor{goldenrod18818934}{RGB}{188,189,34}
\definecolor{gray127}{RGB}{127,127,127}
\definecolor{mediumpurple148103189}{RGB}{148,103,189}
\definecolor{orchid227119194}{RGB}{227,119,194}
\definecolor{sienna1408675}{RGB}{140,86,75}
\definecolor{steelblue31119180}{RGB}{31,119,180}

\pgfplotscreateplotcyclelist{matplotlib}{
steelblue31119180\\
darkorange25512714\\
forestgreen4416044\\
crimson2143940\\
mediumpurple148103189\\
sienna1408675\\
orchid227119194\\
gray127\\
goldenrod18818934\\
darkturquoise23190207\\
steelblue31119180\\
}

\begin{document}

\title[DoNOF]{DoNOF 2.0: A modern Open-Source Electronic Structure Program for Natural Orbital Functionals}

\author{Juan Felipe Huan Lew-Yee}
\email{felipe.lew.yee@quimica.unam.mx}
\affiliation{Donostia International Physics Center (DIPC), 20018 Donostia, Spain.}

\author{Ion Mitxelena}
\email{ion.mitxelena@ehu.eus}
\affiliation{Fisika Aplikatuko departamentua, Vitoria-Gasteiz Ingenieritza Eskola, Euskal Herriko Unibertsitatea (EHU),01006 Vitoria-Gasteiz, Spain}

\author{Jorge M. del Campo}
\email{jmdelc@unam.mx}
\affiliation{Departamento de F\'isica y Qu\'imica Te\'orica, Facultad de Qu\'imica,
Universidad Nacional Aut\'onoma de M\'exico, M\'exico City, C.P. 04510,
M\'exico}

\author{Mario Piris}
\email{mario.piris@ehu.eus}
\affiliation{Donostia International Physics Center (DIPC), 20018 Donostia, Spain.}
\affiliation{Kimika Fakultatea, Euskal Herriko Unibertsitatea (EHU), 20018 Donostia, Spain.}
\affiliation{IKERBASQUE, Basque Foundation for Science, 48013 Bilbao, Spain.}

\date{\today}

\begin{abstract}
In this work, we present the second version of the Donostia Natural Orbital Functional Software, an open-source program for natural orbital functional calculations. The new release incorporates improved optimization algorithms, capabilities for excited-state computations, support for ab initio molecular dynamics, and integration with the \emph{libcint} library. DoNOF~2.0 also extends its property toolbox by enabling the evaluation of nonlinear optical responses, including static polarizabilities and higher-order hyperpolarizabilities via a finite-field Romberg-Richardson scheme.\\

\noindent \textbf{Program Summary} \\
Title: DoNOF \\
Developer’s repository link: http://github.com/DoNOF/ \\
Program's Manual link: https://donof.readthedocs.io/ \\
Licensing provisions: GPLv3 \\
Programming language: Fortran; additional implementations available in Python (PyNOF) and Julia (DoNOF.jl) \\
Multinode capability: Support for distributed execution through a hybrid OpenMPI and OpenMP implementation
\end{abstract}

\maketitle

\section{Introduction}

In recent years, the one-particle reduced density matrix (1RDM) functional theory \cite{Gilbert1975, Levy1979, Valone1980, Schilling2018} has emerged as a promising methodology for investigating challenging chemical systems.\cite{Saubanere2011-oy, Pernal2016-ep, Liebert2022-xz, Ai2022-uh, Cioslowski2023, Sutter2023} In particular, its formulation in the natural orbital (NO) representation, known as the natural orbital functional theory (NOFT), has proven to be the most successful and practical implementation.\cite{Piris2024, Piris2024-zp} The NOFT has shown strong performance in domains traditionally difficult for approximate density functionals, particularly in addressing issues such as charge delocalization error\cite{Lew-Yee2023-ha, Lew-Yee2022-te} and multireference effects.\cite{Piris2011-qn, Ramos-Cordoba2015a, Mitxelena2017a, Mitxelena2018, Mitxelena2020a, Mitxelena2020b, Mitxelena2022-wt, Mitxelena2024-hr, Franco2025-ox} 

The effectiveness of NOFT stems from its ability to incorporate non-dynamic correlation at a reasonable computational cost.\cite{Piris2017-go, Piris2021-sv} Consequently, it allows for the treatment of all orbitals without the need to select a restricted active space, enabling the study of large molecules that are typically inaccessible through standard wavefunction-based approaches. This capability has been demonstrated in recent computations involving iron-porphyrin,\cite{Lew-Yee2023-ke} acenes,\cite{Lew-Yee2025-he} and the metal-to-insulator transition in a system with one thousand hydrogen atoms.\cite{Lew-Yee2025-he, Lew-Yee2025-mh}

NOFT rests on the formal result that the energy is an exact functional of the NOs and their occupation numbers (ONs), although in practice it can only be accessed through approximate forms. Direct variation of this functional is challenging because the reconstructed two-particle reduced density matrix (2RDM) does not necessarily correspond to a valid N-electron state,\cite{Piris2018d} a difficulty known as the N-representability problem of the 2RDM, originally identified by Coleman.\cite{Coleman2007-kq} The N-representability of the 1RDM can, however, be ensured straightforwardly, and its eigenvectors define the NO basis, which offers both practical advantages and valuable chemical insight.\cite{Lowdin1955-qk, Davidson1972-gh, Helbig2010-qo, Piris2013-oo}

In practice, modern NOFs closely resemble Hartree–Fock (HF) and density functional theory (DFT) approaches, as they also require orbital optimization and depend on JK integrals. A key distinction is that a conventional Fockian matrix is not defined and that a constrained optimization of the ONs is also involved. Together, both optimizations (ONs and NOs) are also similar in spirit to multiconfigurational self-consistent field methods, though at a much lower computational cost. As a result, some elements of the machinery behind these methods can be employed, but dedicated specialization is crucial for the carrying out of NOF calculations.\cite{Pernal2005-lz, Piris2009-jo}

The Donostia Natural Orbital Functional (DoNOF) software represents a significant step forward in this direction,\cite{Piris2021-xo} initially supporting single-point calculations that included both pure NOFT and its combination with perturbation theory,\cite{Piris2013c, Piris2018b, Rodriguez-Mayorga2021} together with geometry optimizations\cite{Mitxelena2017, Mitxelena2020c} and thermodynamic analysis for the family of Piris natural orbital functionals (PNOFs).\cite{Piris2013b} Developed in the Basque Country (Spain) the program has enabled collaboration with researchers in Mexico and remains open to contributions worldwide. Since its original release in 2009 as PNOFID,\cite{Piris2009-jo} the software has been strengthened through improvements in stability, reproducibility, and convergence behavior. Moreover, capabilities for direct molecular dynamics\cite{Rivero-Santamaria2024-jd,Piris2024b} and excited-state analysis\cite{Lew-Yee2024-dv} have been incorporated. In addition to the calculation of molecular electric moments,\cite{Mitxelena2016} the program now also introduces the computation of nonlinear optical properties (NLOPs), a new feature presented in this work. These advances motivated the present publication, which introduces the latest version of DoNOF. This paper provides a concise overview of the underlying theory of NOFT and describes the current features of DoNOF, with emphasis on the newly integrated functionalities.

\section{Theory}

For an N-electron quantum system, the state is described by its N-particle density matrix
\begin{equation}
\mathfrak{D}={\displaystyle {\displaystyle {\textstyle {\displaystyle \sum_{i}}}}\omega_{i}\Psi_{i}\left(\mathbf{\mathbf{x}_{\textrm{1}}^{\prime}\mathrm{,}}\ldots\mathbf{\mathrm{,}\mathbf{x}_{\mathrm{N}}^{\prime}}\right)\Psi_{i}^{*}\left(\mathbf{\mathbf{x}_{\textrm{1}}},\ldots,\mathbf{x_{\mathrm{\mathrm{N}}}}\right)}\label{DM}
\end{equation}
where $\omega_{i}$ are positive real numbers with sum one, so that $\mathfrak{D}$ represents a statistical mixture of pure states with weights $\omega_{i}$. Here and throughout, ${\bf x\equiv}\left({\bf r,s}\right)$ denotes the combined spatial and spin coordinates, ${\bf r}$ and ${\bf s}$, respectively.

The electronic energy $E$ for such a system subject to an external potential $v({\bf r})$ is an exact and explicit functional of the 1RDM ($\Gamma$) and the 2RDM ($D$), obtained by contracting $\mathfrak{D}$. It is given by\cite{Piris2007}
\begin{equation}
  E = \sum_{ik} \Gamma_{ik} \; h_{ki} + \sum\limits _{ijkl}D_{ijkl}\left\langle kl|ij\right\rangle
  \label{Energy}
\end{equation}
where $h_{ki}$ and $\braket{kl}{ij}$ denote the usual one- and two-electron integrals in an arbitrary spin-orbital basis $\{ \phi_i(\mathbf{x}) \}$. These are defined, in atomic units, as:
\begin{equation}
h_{ki} =\int d{\bf x} \phi _k ^*({\bf r})\left(-\frac{\nabla ^2 _r} {2} + v({\bf r}) \right) \phi _i ({\bf x})
\end{equation}
\begin{equation}
    \langle kl|ij \rangle= \int \int d{\bf x}_1 d {\bf x}_2 \frac{\phi ^* _k ({\bf x}_1)\phi ^* _l ({\bf x}_2)\phi _i ({\bf x}_1)\phi _j({\bf x}_2)}{|{\bf r}_2 -{\bf r}_1|}
\end{equation}
We adopt Löwdin’s normalization, in which the traces of the 1RDM and 2RDM correspond to the number of electrons and the number of electron pairs, respectively.

The first term in Eq.~(\ref{Energy}) is exactly expressed as a functional of $\Gamma$, whereas the second term, $V_{ee}\left[D\right]$, depends explicitly on the 2RDM. To construct the functional $V_{ee}\left[D\right]$, we work in the NO representation in which the 1RDM is diagonal, $\Gamma_{ik} = n_{i} \delta_{ik}$, so the energy can be expressed as a functional of the ONs and the NOs,
\begin{equation}
  E\left[\left\{ n_{i},\phi_{i}\right\} \right] = \sum\limits _{i}n_{i}h_{ii}+\sum\limits _{ijkl}D[n_{i},n_{j},n_{k},n_{l}]\left\langle kl|ij\right\rangle \label{ENOF}
\end{equation}
Here, $D[n_{i},n_{j},n_{k},n_{l}]$ denotes the reconstructed ensemble 2RDM obtained from the ONs. Restricting the ONs to the interval $0\leq n_{i}\leq1$ provides
a necessary and sufficient condition for ensemble N-representability of the 1RDM.\cite{Coleman1963-ur} However, the functional N-representability problem \cite{Ludena2013} arises because the reconstructed 2RDM must also satisfy its own N-representability conditions.\cite{Mazziotti2012a} While the constraints ensuring an acceptable 1RDM are straightforward to impose, they are not sufficient to guarantee that the resulting 2RDM is N-representable, and therefore not enough to ensure the representability of the approximate functional.

Many necessary conditions for the N-representability of the 2RDM are known, and the problem has been formally solved.\cite{Mazziotti2012} However, a complete set of conditions that does not rely on higher-order RDMs is still unavailable, and a practical solution to the N-representability problem of the 2RDM remains out of reach. What can be done in practice is to impose, in a progressive manner, necessary N-representability conditions on the 2RDM in order to obtain an approximate form that leads to an energy with a physically meaningful character, that is, one that approximates the behavior of a valid N-electron density matrix. This bottom-up strategy\cite{Piris2013e} generates a set of inequalities that has led to the family of functionals known in the literature as Piris NOFs (PNOFs)\cite{Piris2013b} and Global NOFs (GNOFs).\cite{Piris2021-sv, Lew-Yee2025-he}

Consider now the cumulant expansion of the 2RDM in the NO representation,\cite{Mazziotti1998}
\begin{equation}
D_{ijkl}=\frac{n_{i}n_{j}}{2}\left(\delta_{ik}\delta_{jl}-\delta_{il}\delta_{jk}\right)+\lambda_{ijkl}
\end{equation}
Here, the 2RDM is decomposed into an antisymmetric product of 1RDMs, which is simply the HF approximation, and a correction term $\lambda$. The latter is known as the cumulant or correlation matrix.\cite{Kutzelnigg1999-xj}

We restrict ourselves to an N-electron Hamiltonian that does not depend on spin coordinates, so both $\widehat{S}_{z}$ and $\widehat{S}^{2}$ commute with it. For eigenstates of $\widehat{S}_{z}$, only those 2RDM blocks that conserve the number of electrons of each spin type are non-vanishing, that is, $D^{\alpha\alpha}$, $D^{\beta\beta}$, and $D^{\alpha\beta}$. 
The use of the (2,2)-positivity N-representability conditions \cite{Mazziotti2012} for reconstructing $\lambda$ was proposed in Ref.\citenum{Piris2006}. This particular reconstruction is based on the introduction of two auxiliary matrices $\Delta$ and $\Pi$ expressed in terms of the ONs. In a spin restricted formulation, $\left\{ \left|i\right\rangle \right\} =\left\{ \left|p\sigma\right\rangle \right\} $, the structure of the two-particle cumulant is
\begin{eqnarray}    
\lambda_{pq,rt}^{\sigma\sigma}=-\frac{\Delta_{pq}}{2}\left(\delta_{pr}\delta_{qt}-\delta_{pt}\delta_{qr}\right)\; , \;\;\; \sigma=\alpha,\beta\\
\lambda_{pq,rt}^{\alpha\beta}=-\frac{\Delta_{pq}}{2}\delta_{pr}\delta_{qt}+\frac{\Pi_{pr}}{2}\delta_{pq}\delta_{rt} \;\;\;\;\;\;\;\;\;\;\;\;\;\;\;\;\;\;
\end{eqnarray}

\begin{figure}
    \centering
    \begin{tikzpicture}

    \node[align=center,rotate=90] at (-1.5,10) {\small{Weakly Double} \\ \small{Occupied Orbitals} \\ \small{$\Omega_\mathrm{II}$}};
    \node[align=center,rotate=90] at (-1.5,5) {\small{Single} \\ \small{Occupied Orbitals} \\ \small{$\Omega_\mathrm{I}$}};
    \node[align=center,rotate=90] at (-1.5,2) {\small{Strongly Double} \\ \small{Occupied Orbitals} \\ \small{$\Omega_\mathrm{II}$}};

    \node[align=right] at (-0.5,3.9) {\small{$\mathrm{N_{II}}/2$}};
    \node[align=right] at (-0.5,6.4) {\small{$\phantom{/2}\mathrm{N_\Omega}$}};
    
    \draw[color1,->,very thick] (3.5,1) -- ++(1,0) -- ++(0,8) -- ++(-1,0) ;
    \draw[color1,->,very thick](3.5,1) -- ++(1,0) -- ++(0,11) -- ++(-1,0);
    \node[color1,anchor=west] at (3.25,1.25) {$\Omega_1$};

    \draw[color2,->,very thick] (3.5,2) -- ++(0.75,0) -- ++(0,6) -- ++(-0.75,0) ;
    \draw[color2,->,very thick](3.5,2) -- ++(0.75,0) -- ++(0,9) -- ++(-0.75,0);
    \node[color2,anchor=west] at (3.25,2.25) {$\Omega_2$};

    \draw[color3,->,very thick] (3.5,3) -- ++(0.5,0) -- ++(0,4) -- ++(-0.5,0) ;
    \draw[color3,->,very thick](3.5,3) -- ++(0.5,0) -- ++(0,7) -- ++(-0.5,0);
    \node[color3,anchor=west] at (3.25,3.25) {$\Omega_3$};



    \draw [color1,very thick] (0.3,12) -- (2.75,12);
    \draw[->,very thick] (1,12) -- (1,12.125);
    \draw[<-,very thick] (2,12) -- (2,12.125);

    \draw [color2,very thick] (0.3,11) -- (2.75,11);
    \draw[->,very thick] (1,11) -- (1,11.15);
    \draw[<-,very thick] (2,11) -- (2,11.15);

    \draw [color3,very thick] (0.3,10) -- (2.75,10);
    \draw[->,very thick] (1,10) -- (1,10.175);
    \draw[<-,very thick] (2,10) -- (2,10.175);



    \draw [color1,very thick] (0.3,9) -- (2.75,9);
    \draw[->,very thick] (1,9) -- (1,9.2);
    \draw[<-,very thick] (2,9) -- (2,9.2);

    \draw [color2,very thick] (0.3,8) -- (2.75,8);
    \draw[->,very thick] (1,8) -- (1,8.225);
    \draw[<-,very thick] (2,8) -- (2,8.225);

    \draw [color3,very thick] (0.3,7) -- (2.75,7);
    \draw[->,very thick] (1,7) -- (1,7.25);
    \draw[<-,very thick] (2,7) -- (2,7.25);


    \draw[dashed, thick] (0,6.4) -- (3,6.4);


    \draw [black,very thick] (0.3,5.5) -- (2.75,5.5);
    \draw[->,very thick] (1,5.5) -- (1,5.8);
    \draw[<-,very thick] (2,5.5) -- (2,5.8);

    \draw [black,very thick] (0.3,4.5) -- (2.75,4.5);
    \draw[->,very thick] (1,4.5) -- (1,4.8);
    \draw[<-,very thick] (2,4.5) -- (2,4.8);

    \draw[dashed, thick] (0,3.9) -- (3,3.9);


    \draw [color3,very thick] (0.3,3) -- (2.75,3);
    \draw[->,very thick] (1,3) -- (1,3.45);
    \draw[<-,very thick] (2,3) -- (2,3.45);

    \draw [color2,very thick] (0.3,2) -- (2.75,2);
    \draw[->,very thick] (1,2) -- (1,2.475);
    \draw[<-,very thick] (2,2) -- (2,2.475);

    \draw [color1,very thick] (0.3,1) -- (2.75,1);
    \draw[->,very thick] (1,1) -- (1,1.5);
    \draw[<-,very thick] (2,1) -- (2,1.5);

\end{tikzpicture}
    \caption{Representation of the orbital-pairing structure for a system with 8 electrons. In this example, $S = 1$ (triplet) and $\mathrm{N_{I}} = 2$, so two orbitals make up the subspace $\Omega_\text{I}$, whereas six electrons ($\mathrm{N_{II}} = 6$) make up the subspace $\Omega_\text{II}$. Note that $\mathrm{N}_{g} = 2$. The arrows depict the values of the ensemble ONs, $\alpha$ ($\uparrow$) or $\beta$ ($\downarrow$), in each orbital.}
    \label{fig:pairing}
\end{figure}
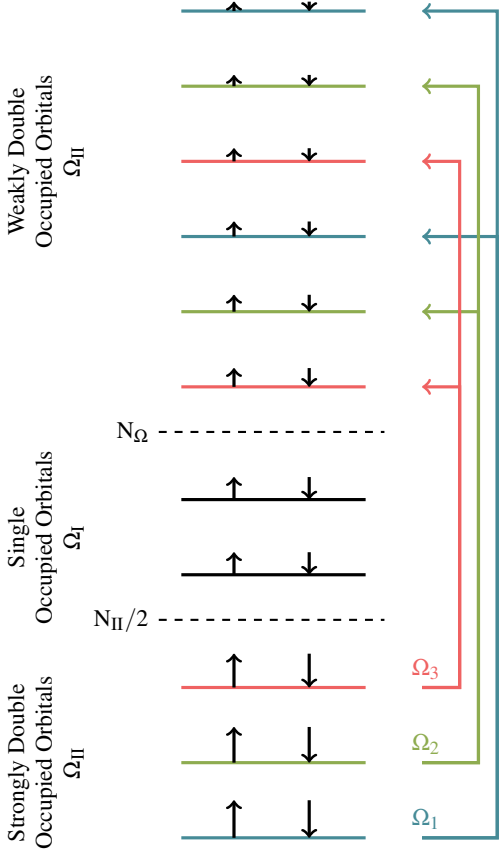

The main advantage of these cumulant reconstructions is that they lead to a JKL-only energy expression, namely,
\begin{eqnarray}
 E = 2 \sum_p n_p h_{pp} + \sum_{pq} A[n_p,n_q] J_{qp} \;\;\;\;\;\;\;\;\; \nonumber \\ 
  - \sum_{pq} B[n_p,n_q] K_{qp} - \sum_{pq} C[n_p,n_q] L_{qp} 
 \label{PNOF}
\end{eqnarray}
where $J_{pq}$, $K_{pq}$ and $L_{pq}$ denote the Coulomb, exchange and exchange-time-inversion integrals,\cite{Piris1999} respectively, and where $A$, $B$ and $C$ are coefficients that depend on the ONs. For explicit formulations, we refer the reader to the corresponding references. In this work, we consider the cumulant reconstructions underlying the electron-pairing models\cite{Piris2018e} PNOF5,\cite{Piris2011-qn, Piris2013e} PNOF7,\cite{Piris2017-go, Mitxelena2018a} GNOF,\cite{Piris2021-sv} and GNOFm,\cite{Lew-Yee2025-he} whose orbital-pairing structure is shown in Fig.~\ref{fig:pairing}.\cite{BOUTOU2025}

It is assumed that $\mathrm{N_{I}}$ unpaired electrons determine the total spin $S$ of the system, while the remaining electrons ($\mathrm{N_{II}}=\mathrm{N-N_{I}}$) form spin pairs, so that their contributions to the total spin cancel out. We focus on the mixed state of highest multiplicity, $2S+1=\mathrm{N_{I}}+1$, with $S=\mathrm{N_{I}}/2$. It is worth noting that the expectation value of $\hat{S}_{z}$ for the ensemble as a whole is zero.\cite{Piris2019} 

Accordingly, the orbital space is partitioned as $\Omega=\Omega_\text{I} \oplus \Omega_\text{II}$. The subspace $\Omega_{\mathrm{II}}$ consists of $\mathrm{N_{II}}/2$ mutually disjoint subspaces $\Omega{}_{g}$. Each $\Omega{}_{g}\in\Omega_{\mathrm{II}}$ contains one strongly doubly occupied orbital $\left|g\right\rangle $ with $g\leq\mathrm{N_{II}}/2$, and $\mathrm{N}_{g}$ weakly doubly occupied orbitals $\left|p\right\rangle $ with $p>\mathrm{\mathrm{N}_{\Omega}}$, where $\mathrm{\mathrm{N}_{\Omega}=}\mathrm{N_{II}}/2+\mathrm{N_{I}}$. These subspaces correspond to the sets of orbitals depicted with the same color in Fig.~\ref{fig:pairing}. That is,
\begin{equation}
\Omega{}_{g}=\left\{ \left|g\right\rangle ,\left|p_{1}\right\rangle ,\left|p_{2}\right\rangle ,...,\left|p_{\mathrm{N}_{g}}\right\rangle \right\} \label{OmegaG}
\end{equation}

Taking into account the spin, the total occupancy for a given subspace $\Omega{}_{g}$ is 2, which is reflected in the following sum rule:
\begin{equation}
n_{g}+\sum_{i=1}^{\mathrm{N}_{g}}n_{p_{i}}=1,\quad g=1,2,...,\frac{\mathrm{N_{II}}}{2}\label{sum1}
\end{equation}

In general, $\mathrm{N}_{g}$ may differ from one subspace to another, but it must be large enough to describe each electron pair adequately. In our implementation, $\mathrm{N}_{g}$ is taken as a fixed value for all subspaces $\Omega_{g}\in\Omega_{\mathrm{II}}$. The maximum admissible value of $\mathrm{N}_{g}$ is determined by the basis set employed in the calculations.

For multiplets, $\Omega_{\mathrm{I}}$ is composed of $\mathrm{N_{I}}$ mutually disjoint subspaces $\Omega{}_{g}$. In contrast to $\Omega_{\mathrm{II}}$, each subspace $\Omega_{g} \in \Omega_{\mathrm{I}}$ contains only one orbital $g$ with $n_{g}=1/2$. Each of these orbitals is therefore singly occupied, although the spin of the unpaired electron ($\alpha$ or $\beta$) is not specified. 

It is then clear that the sum of the ONs of all orbitals in $\Omega_{\mathrm{I}}$ equals $\mathrm{N_{I}}$. Eq.~(\ref{sum1}) further implies that the ONs in $\Omega_{\mathrm{II}}$ add up to $\mathrm{N_{II}}$. Consequently, the trace of the 1RDM is equal to the total number N of electrons.\cite{Piris2019}

Minimizing Eq.~(\ref{PNOF}) leads to a demanding constrained optimization problem. In practice, the energy is optimized separately with respect to ONs and NOs, as simultaneous optimization has so far proven to be inefficient. This separation allows the ONs to be fully optimized at reasonable cost while focusing computational effort on the more challenging NO optimization. Indeed, the ONs typically converge to near-final values within the first few outer iterations, leaving the optimization of the NOs as the dominant task.

\subsection{Occupation Number Optimization\label{sec:ON-opt}}

Direct optimization of ONs under both the 1RDM N-representability constraints and the pairing conditions~(\ref{sum1}) leads to an impractical constrained minimization problem. A more effective strategy is to express the set of occupation numbers $\{n_p\}$ in terms of a new set of auxiliary variables $\{\gamma_p\}$ chosen so that these constraints are automatically satisfied. This transformation enables an unconstrained optimization with respect to the $\{\gamma_p\}$ variables. In DoNOF, two different transformations of this type are available.

\textbf{Trigonometric mapping} of the ONs exploits the Pythagorean trigonometric identity to ensure compliance with Eq.~(\ref{sum1}). Accordingly, the ON of the strongly doubly occupied orbital $g$ is given by
\begin{equation}
    n_g = \frac{1}{2} \left(1 + \cos^2 \gamma_g \right),
    \label{eq:trigo-n_sdono_gamma}
\end{equation}
which naturally defines the corresponding hole $h_g = 1-n_g$. The ONs of the weakly occupied orbitals are then obtained by successive multiplication of the cumulative hole with a squared sine function,
\begin{eqnarray}
    n_{p_1} &=& h_g \sin^2 \gamma_{p_1} ,\nonumber\\
    n_{p_2} &=& h_g \cos^2 \gamma_{p_1} \sin^2 \gamma_{p_2},\nonumber\\
    &\vdots& \nonumber\\
    n_{p_{i}} &=& h_g \cos^2 \gamma_{p_1} \cos^2 \gamma_{p_2} \cdots \cos^2 \gamma_{p_{i-1}} \sin^2 \gamma_{p_i}, \\
    &\vdots& \nonumber\\    
    \nonumber
\end{eqnarray}
\begin{eqnarray}
   n_{p_\mathrm{N_{g}-1}} &=& h_g \cos^2 \gamma_{p_1} \cos^2 \gamma_{p_2} \cdots \cos^2 \gamma_{p_{N_{g}-2}} \sin^2 \gamma_{p_\mathrm{N_{g}-1}}, \nonumber\\    
   n_{p_\mathrm{N_{g}}} &=& h_g \cos^2 \gamma_{p_1} \cos^2 \gamma_{p_2} \cdots \cos^2 \gamma_{p_{N_{g}-2}} \cos^2 \gamma_{p_\mathrm{N_{g}-1}} \nonumber\\ 
   \nonumber
\end{eqnarray}

\textbf{Softmax mapping} of the ONs makes use of the generalized logistic (softmax) function, so that the strongly doubly occupied orbital is given by 
\begin{equation}
    n_g = \frac{1}{1 + \sum_{k=1}^{N_{g}}\exp(\gamma_{p_k})} ,
    \label{eq:softmax-n_sdono_gamma}
\end{equation}
whereas the ONs of the weakly occupied orbitals $n_{p_i}$ are obtained by multiplying $n_g$ by suitable exponential functions of the corresponding $\gamma_{p_i}$ variables:
\begin{eqnarray}
    \label{eq:n_wdono_gamma1}
    n_{p_1} &=& n_g \exp(\gamma_{p_1}) ,\nonumber\\
    &\vdots& \nonumber\\
    n_{p_i} &=& n_g \exp(\gamma_{p_i}) ,\\
   &\vdots& \nonumber\\  
    n_{p_\mathrm{N_{g}}} &=& n_g \exp(\gamma_{p_{N_g}}) ,\nonumber\\
     \nonumber
\end{eqnarray}
This mapping is very similar to the one reported in Ref.~\citenum{Franco2024}, except that one auxiliary variable has been removed.

In both trigonometric and softmax transformations, one fewer auxiliary variable than the number of orbitals in the subspace is required, reflecting the fact that a given ON can be obtained from the remaining ones due to the pairing condition; therefore, one degree of freedom is effectively eliminated by the mappings. In most cases, the softmax transformation is numerically more stable. However, Eq.~(\ref{eq:softmax-n_sdono_gamma}) does not guarantee that $n_g$ is the largest ON nor that it is at least one half. The former issue can be addressed by sorting the orbitals, whereas the latter is guaranteed only by Eq.~(\ref{eq:trigo-n_sdono_gamma}). Nevertheless, this limitation is generally not problematic, except in specific situations.
 
\subsection{Natural Orbital Optimization\label{sec:NO-opt}}

For fixed ONs, optimizing NOs under orthonormality constraints can be performed using the Lagrange multiplier technique as follows:
\begin{equation}
    \mathcal{L} \left[\left\{\varphi_{p}\right\}\right] = E\left[\left\{\varphi_{p}\right\}\right] - 2\sum_{pq} \lambda_{qp} (\braket{\varphi_{p}}{\varphi_{q}} - \delta_{pq})
\end{equation}
In the standard self-consistent field approach, orbital optimization is performed in the canonical representation, where $\lambda$ is diagonal. However, $\Gamma$ and $\lambda$ cannot be diagonalized simultaneously, except in the HF case, which prevents the construction of a true Fockian in NOF calculations.\cite{Piris2013-oo} This limitation can be mitigated by exploiting the symmetry of $\lambda$ at the minimum to build a generalized pseudo-Fockian with well-defined off-diagonal elements, leading to the iterative diagonalization method.\cite{Piris2009-jo} This procedure is implemented in DoNOF as an optional orbital-optimization scheme. Although this represented a major step forward,\cite{Piris2021-xo} its efficiency remains limited by the absence of an exact expression for the diagonal elements and by the lack of effective acceleration techniques.

A well-established alternative to diagonalization is the orbital-rotation approach. In DoNOF, this approach is used, whereby the updated orbitals are obtained through a unitary transformation of the previous ones,
\begin{equation}
    \varphi_p^{(i+1)}(\mathbf{r}) = \sum_{q} U_{qp} \varphi_q^{(i)}(\mathbf{r})
\end{equation}
where $\textbf{U}$ is a unitary matrix expressed as the exponential of an antisymmetric matrix $\textbf{y}$ ($y_{pq} = -y_{qp}$)
\begin{equation}
    \textbf{U} = \exp(\textbf{y})
\end{equation}
In matrix form, this becomes
\begin{equation}
    \textbf{C}^{(i+1)} = \textbf{U} \textbf{C}^{(i)}    
\end{equation}
where $\textbf{C}^{(i)}$ denotes the NO coefficient matrix at the i-iteration. Orbital optimization can therefore be performed by minimizing the energy with respect to the independent variables $y_{pq}$ without constraints. Trust-region methods are commonly used for this purpose, but their reliance on Hessian\cite{mitxelena2018b} evaluations increases computational cost.

For this reason, we disregard the Hessian and perform orbital optimization using only the orbital gradient, which can be obtained from the matrix of Lagrange multipliers as
\begin{eqnarray}
    g^{(i)}_{pq} = \frac{dE^{(i)}}{dy_{pq}} = 4(\lambda^{(i)}_{pq} - \lambda^{(i)}_{qp})
\end{eqnarray}
Inspired by techniques from deep learning, we employ the adaptive-momentum (ADAM) algorithm to optimize the NOs. This method updates the variables using first and second moments computed solely from orbital gradients, providing efficient scaling and very low memory requirements. Details of our ADAM implementation for NO optimization, as well as its integration with ON optimization, are provided in Ref.~\citenum{Lew-Yee2025-he}.

\section{Computational Details}

DoNOF is written in Fortran and compiled with \emph{gfortran}, with parallel implementations for both single-node (OpenMP) and multi-node (MPI) execution. Particular attention has been given to the use of external libraries for tasks not specific to NOFT. Accordingly, the \emph{LAPACK} library is employed for linear algebra routines, while the \emph{libcint} library is used to compute one- and two-electron integrals.\cite{Sun2015} This enables calculations to be performed using either cartesian or spherical Gaussian basis functions.

In DoNOF, the input file specifies all relevant variables for a given calculation through two fundamental namelists, \texttt{INPRUN} and \texttt{NOFINP}, which allow the user to provide all required data in free-format style. These namelists define the parameters that control both the general type of calculation and the details of the functional optimization procedure. A complete description of all variables included in these namelists can be found in the online manual available at https://donof.readthedocs.io/. The type of calculation to be performed in DoNOF is governed by the keyword \texttt{RUNTYP}, which may take the values \texttt{`ENERGY'}, \texttt{`GRAD'}, \texttt{`HESS'}, \texttt{`OPTGEO'}, and \texttt{`DYN'}, corresponding respectively to a single-point energy calculation, computation of energies and analytic gradients with respect to nuclear coordinates, numerical Hessian evaluation from analytic gradients, molecular geometry optimization, and Born–Oppenheimer on-the-fly molecular dynamics. In the following, we briefly discuss these calculation options.

\section{Single Point Energy Calculations}

Single-point calculations can be performed using any of the available functionals. A typical input file is shown below:

\begin{verbatim}
 &INPRUN RUNTYP=`ENERGY' MULT=1 ICHARG=0 /
 $DATA
 Title: Example N2
 cc-pVDZ
 N  7.0  0.0000  0.0000 -0.5488
 N  7.0  0.0000  0.0000  0.5488
 $END
 &NOFINP /
\end{verbatim}

As illustrated by the example, in addition to defining the type of calculation, \texttt{INPRUN} controls essential system parameters such as the multiplicity and the total charge. Although their default values are \texttt{MULT=1} and \texttt{ICHARG=0}, respectively, these parameters are typically specified explicitly to avoid ambiguity. The remaining keywords include predefined default options. For instance, one may choose to compute the four-center electron repulsion integrals exactly by setting \texttt{ERITYP=`FULL'}, which scales as fifth order, or rely on the resolution of the identity using the default option \texttt{ERITYP=`RI'}, which reduces the cost of the orbital transformation to fourth-order scaling. Another important keyword is \texttt{GTYP}, which can be set to \texttt{`CART'} or \texttt{`SPH'} (default) to select between cartesian and spherical Gaussian basis functions, respectively.

In the example, one can also see the \texttt{NOFINP} namelist, where all variables have predefined default values, allowing the optimization procedure to be handled automatically and enabling the code to function as a true black box. The detailed behavior of the optimization can be adjusted here. For instance, the keyword \texttt{ISOFTMAX} =$\{$0,1$\}$ determines how the ONs are transformed to fulfill the N-representability and pairing constraints, as described in subsection~\ref{sec:ON-opt}. Choosing 1 applies the softmax transformation, whereas 0 employs a trigonometric transformation. Another essential keyword is \texttt{IORBOPT}=$\{$1,2$\}$, settled to 1 enables the iterative diagonalization scheme, while setting it to 2 selects the ADAM-based orbital optimization using orbital rotations, as discussed in subsection~\ref{sec:NO-opt}.

The functional to be used is specified through the \texttt{IPNOF} keyword, which by default is set to 8 at the time of writing this article. Currently, DoNOF includes seven PNOFs and a global functional (GNOF), along with specific variants of some of them. The most commonly used options are listed in Table~\ref{tab:ipnof}.

\begin{table}[htbp]
    \caption{Common NOF options in DoNOF, selected through the \texttt{IPNOF} and \texttt{Imod} keywords.}
    \bigskip
    \centering
    \begin{tabular}{l|cc}
    NOF & \multicolumn{2}{c}{Keywords} \\
    \hline
    PNOF5  & \multicolumn{2}{c}{IPNOF=5} \\
    PNOF7  & \multicolumn{2}{c}{IPNOF=7} \\
    GNOF   & \; IPNOF=8 & Imod=0 \\
    GNOFm  & \; IPNOF=8 & Imod=1 \\
    \end{tabular}
    \label{tab:ipnof}
\end{table}

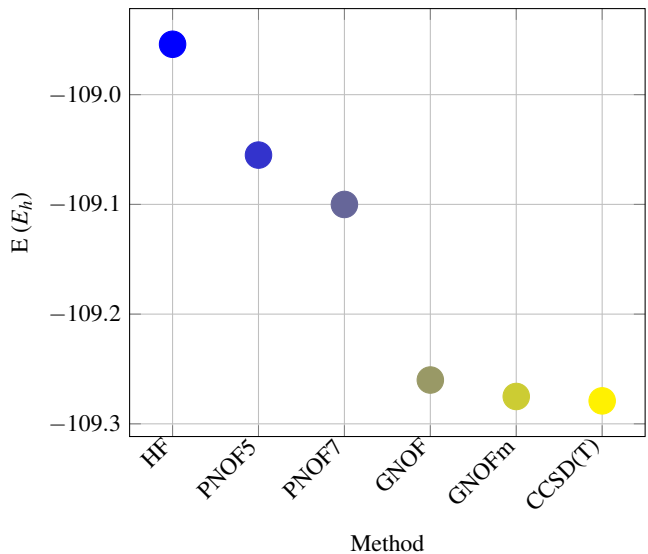
\begin{figure}[htbp]
    \centering
    \begin{tikzpicture}
    \definecolor{cividis1}{rgb}{0.00, 0.13, 0.30}
    \definecolor{cividis2}{rgb}{0.13, 0.30, 0.50}
    \definecolor{cividis3}{rgb}{0.33, 0.46, 0.62}
    \definecolor{cividis4}{rgb}{0.55, 0.63, 0.68}
    \definecolor{cividis5}{rgb}{0.78, 0.81, 0.73}
    \definecolor{cividis6}{rgb}{1.00, 1.00, 0.92}

    \begin{axis}[
        xlabel={Method},
        ylabel={E ($E_h$)},
        symbolic x coords={HF,PNOF5,PNOF7,GNOF,GNOFm,CCSD(T)},
        x tick label style={rotate=45, anchor=east},
        grid=major,
        y tick label style={/pgf/number format/.cd,fixed,fixed zerofill,precision=1},
    ]
    
    \addplot[only marks, mark=*, mark size=5, blue] coordinates {(HF, -108.954)};
    \addplot[only marks, mark=*, mark size=5, blue!80!yellow] coordinates {(PNOF5, -109.055)};
    \addplot[only marks, mark=*, mark size=5, blue!60!yellow] coordinates {(PNOF7, -109.100)};
    \addplot[only marks, mark=*, mark size=5, blue!40!yellow] coordinates {(GNOF, -109.260)};
    \addplot[only marks, mark=*, mark size=5, blue!20!yellow] coordinates {(GNOFm, -109.275)};
    \addplot[only marks, mark=*, mark size=5, yellow] coordinates {(CCSD(T), -109.279)};
    \end{axis}
    \end{tikzpicture}
    \caption{Energy of \ce{N2}/cc-pVDZ computed using several NOFs, compared with HF and CCSD(T).}
    \label{fig:energy-nofs-n2}
\end{figure}

Fig.~\ref{fig:energy-nofs-n2} illustrates the performance of the most commonly used NOFs by comparing their energies with the HF and CCSD(T) reference values for the triple-bonded \ce{N2} molecule at its equilibrium geometry, using Dunning’s cc-pVDZ basis set.\cite{Dunning} Since all these functionals employ the electron-pairing approach,\cite{Piris2018e} the total electron correlation separates naturally into intra-pair and inter-pair components. PNOF5 corresponds to an independent-pair model,\cite{Piris2011-qn, Piris2013e} in which correlation is restricted to orbitals within the same pair. This leads to a smaller amount of total electron correlation; nevertheless, this intra-pair component is essential for describing the nature of chemical bonds,\cite{Matxain2012a} and it plays a crucial role in bond-breaking processes\cite{Ruiperez2013} and radical-formation reactions.\cite{Lopez2012}

PNOF7 is the first functional to incorporate correlation between orbitals belonging to different pairs,\cite{Piris2017-go, Mitxelena2018a} thereby accounting for static inter-pair effects. Both PNOF5 and PNOF7 lack dynamic inter-pair correlation, which can be recovered in DoNOF through the perturbative corrections activated by the logical keywords \texttt{SC2MCPT}, \texttt{OIMP2}, and \texttt{MBPT}, as implemented in our code.\cite{Piris2013c, Piris2018b, Rodriguez-Mayorga2021} GNOF addresses electron correlation using a global strategy,\cite{Piris2021-sv} while GNOFm refines this description by extending the set of correlated orbitals to include those lying below the Fermi level, thereby recovering additional dynamic correlation.\cite{Lew-Yee2025-he} It is important to emphasize that understanding complex chemical reactivity requires focusing on chemical behavior rather than absolute energies; nevertheless, Fig.~\ref{fig:energy-nofs-n2} shows how the functionals progressively approach the absolute energies provided by the gold-standard CCSD(T) reference.

In single-point energy calculations, DoNOF provides access to a variety of molecular properties through specific options in the \texttt{NOFINP} namelist. These include Mulliken population analysis (\texttt{IMULPOP=1}); diagonalization of the Lagrange-multiplier matrix to obtain canonical molecular orbitals and their corresponding one-particle energies (\texttt{DIAGLAG=T}); and the generation of several standard output formats for post-processing and visualization, such as AIMPAC wavefunction files (\texttt{IAIMPAC=1}), formatted-checkpoint files (\texttt{IFCHK=1}), and MOLDEN input files (\texttt{IMOLDEN=1}). Additional capabilities include the construction of an initial-conditions file (\texttt{ini.xyz}) based on normal modes and zero-point energy velocities (\texttt{INICOND=1}), as well as the printing of atomic reduced density matrices (\texttt{NOUTRDM=1}). Building upon these established features, the current version of DoNOF also incorporates two major extensions to single-point functionality: the calculation of excited states and the evaluation of electric response properties, including dipole polarizabilities and nonlinear optical hyperpolarizabilities. These enhancements significantly broaden the scope of NOF-based electronic-structure analysis.

\subsection{Excited States}

DoNOF 2.0 provides access to both charged and neutral excitations. These two classes of excited states differ only in the excitation operator used to construct them, while in both cases the coefficients defining the excited states are obtained directly from the ground-state first- and second-order RDMs. As a result, excited-state properties can be evaluated without performing separate orbital optimizations, offering an efficient and internally consistent route to excitation energies within the NOFT framework.

Charged excitations are computed using the extended Koopmans’ theorem (EKT),\cite{Day1974} which relates the 1RDM and 2RDM of a Coulombic system to its ionization potentials (IPs) and electron affinities (EAs). In DoNOF, IPs are activated by setting \texttt{IEKT=1} in the \texttt{NOFINP} namelist. When combined with NOF approximations, EKT generally provides reliable values for the lowest IPs,\cite{Piris2012} although its accuracy decreases for higher ionization energies. Direct computation of EAs via EKT is considerably more problematic and currently yields unsatisfactory results. A more practical alternative is to approximate the EA as the negative of the IP of the corresponding anionic species, computed at the experimental geometry of the neutral molecule. This strategy improves the agreement with experimental values and performs better than the conventional Koopmans’ theorem.\cite{Piris2012}

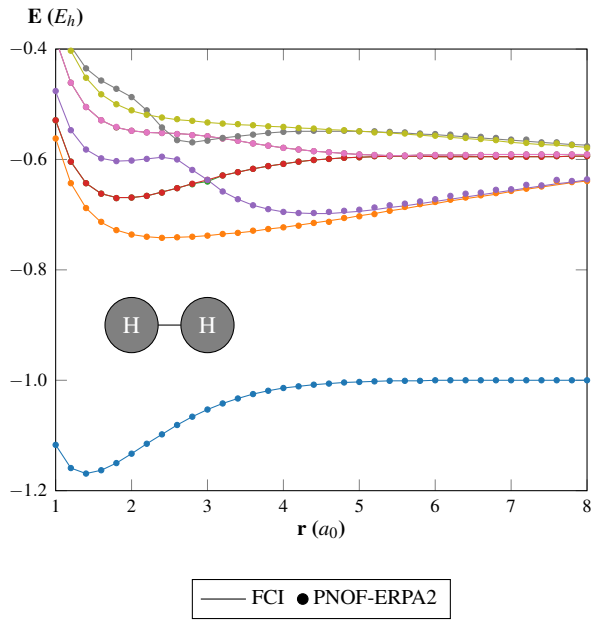
\begin{figure}
    \centering
    \begin{tikzpicture}
\begin{axis}[
    width=\linewidth,
    enlargelimits=false,
    cycle list name = matplotlib,
    xmin = 1.0,
    xmax = 8.0,
    ymax = -0.4,
    ymin = -1.2,
    xlabel = r ($a_0$),
    ylabel = E ($E_h$),
    y tick label style={
        /pgf/number format/.cd,
        fixed,
        fixed zerofill,
        precision=1,
        /tikz/.cd
    },
    every axis y label/.style={
    at={(ticklabel* cs:1.05)},
    anchor=south,
    },
    legend style={/tikz/every even column/.append style={column sep=0.1cm},at={(0.5,-0.2)},anchor=north,font=\footnotesize},
    legend columns=2,
    ylabel style={align=center, inner sep=0pt, font=\footnotesize\bfseries\boldmath},
    xlabel style={align=center, inner sep=0pt, font=\footnotesize\bfseries\boldmath},
    x tick label style={font=\scriptsize\bfseries\boldmath},
    y tick label style={font=\scriptsize\bfseries\boldmath},
]

\addplot+[steelblue31119180,forget plot] table[x=r, y=E0] {h2-fci.dat};
\addplot+[darkorange25512714,forget plot] table[x=r, y=E1] {h2-fci.dat};
\addplot+[forestgreen4416044,forget plot] table[x=r, y=E2] {h2-fci.dat};
\addplot+[crimson2143940,forget plot] table[x=r, y=E3] {h2-fci.dat};
\addplot+[mediumpurple148103189,forget plot] table[x=r, y=E4] {h2-fci.dat};
\addplot+[sienna1408675,forget plot] table[x=r, y=E5] {h2-fci.dat};
\addplot+[orchid227119194,forget plot] table[x=r, y=E6] {h2-fci.dat};
\addplot+[gray127,forget plot] table[x=r, y=E7] {h2-fci.dat};
\addplot+[goldenrod18818934,forget plot] table[x=r, y=E8] {h2-fci.dat};

\addplot[] coordinates {(1,1)(2,2)};
\addlegendentry{FCI}

\addplot+[steelblue31119180,only marks,mark size=1pt,forget plot] table[x=r, y=E0] {h2-pnof-erpa2.dat};
\addplot+[darkorange25512714,only marks,mark size=1pt,forget plot] table[x=r, y=E1] {h2-pnof-erpa2.dat};
\addplot+[forestgreen4416044,only marks,mark size=1pt,forget plot] table[x=r, y=E2] {h2-pnof-erpa2.dat};
\addplot+[crimson2143940,only marks,mark size=1pt,forget plot] table[x=r, y=E3] {h2-pnof-erpa2.dat};
\addplot+[mediumpurple148103189,only marks,mark size=1pt,forget plot] table[x=r, y=E4] {h2-pnof-erpa2.dat};
\addplot+[sienna1408675,only marks,mark size=1pt,forget plot] table[x=r, y=E5] {h2-pnof-erpa2.dat};
\addplot+[orchid227119194,only marks,mark size=1pt,forget plot] table[x=r, y=E6] {h2-pnof-erpa2.dat};
\addplot+[gray127,only marks,mark size=1pt,forget plot] table[x=r, y=E7] {h2-pnof-erpa2.dat};
\addplot+[goldenrod18818934,only marks,mark size=1pt,forget plot] table[x=r, y=E8] {h2-pnof-erpa2.dat};

\addplot[only marks,mark size=2pt] coordinates {(1,1)(2,2)};
\addlegendentry{PNOF-ERPA2}

\draw[fill=gray!100,text=white] (2.0,-0.9) ellipse (0.35 and 0.05) node {H};
\draw (2.35,-0.9) -- (2.65,-0.9);
\draw[fill=gray!100,text=white] (3.0,-0.9) ellipse (0.35 and 0.05) node {H};

\end{axis}
\end{tikzpicture}
    \caption{Potential energy curve of \ce{H2} with excited states computed using PNOF-ERPA2 and the cc-pVDZ basis set.}
    \label{fig:h2-PNOF-ERPA}
\end{figure}

Neutral excitations are computed using the extended random phase approximation (ERPA), with three levels of approximation: single non-diagonal excitations (ERPA0), additional single diagonal excitations (ERPA1), and additional double diagonal excitations (ERPA2).\cite{Chatterjee2012-qn,Pastorczak2015-wz} ERPA is applied directly to the matrices resulting from the NOF ground-state calculation, leading to equations analogous to those encountered in time-dependent density functional theory (TD-DFT). These calculations are activated by setting \texttt{ERPA=T} in the input, as illustrated below:
\begin{verbatim}
 &INPRUN RUNTYP=`ENERGY' MULT=1 ICHARG=0 /
 $DATA
 Title: Excited States H2
 cc-pVDZ
 H  1.0 -0.3707  0.0000  0.0000
 H  1.0  0.3707  0.0000  0.0000
 $END
 &NOFINP IPNOF=8 Imod=1 ERPA=T /
\end{verbatim}
A key advantage of this approach is its ability to incorporate the ONs and the inherent multireference character of the ground state.\cite{Lew-Yee2024-dv} Fig.~\ref{fig:h2-PNOF-ERPA} illustrates this for the potential energy curve of \ce{H2}. The blue curve represents the PNOF ground-state energy (all PNOFs coincide for a two-electron system), while the remaining curves show the excitation energies added to this reference. The PNOF-ERPA method, shown with circular markers, reproduces the Full-CI curve with excellent accuracy, including in the dissociation region. In practice, ERPA0 and ERPA1 often provide sufficiently accurate results at lower computational cost, whereas more efficient strategies for solving the ERPA2 equations remain an active area of research.

\subsection{NLOPs: Polarizabilities and Hyperpolarizabilities}

DoNOF~2.0 incorporates the capability to compute static polarizabilities ($\alpha$) and first ($\beta$) and second ($\gamma$) hyperpolarizabilities by evaluating the dipole-moment response to an external electric field. The calculations are activated through the keyword \texttt{NLOP} in the \texttt{INPRUN} namelist, which specifies whether $\alpha$ (\texttt{NLOP=1}), $\beta$ (\texttt{NLOP=2}), $\gamma$ (\texttt{NLOP=3}) or all three properties (\texttt{NLOP=-1}) are evaluated. Additional control parameters allow the user to define the number of field points (\texttt{NPOINT}) and the initial step size (\texttt{STEP}), which together determine the dyadically scaled sequence of electric-field strengths used in the numerical differentiation.

Before detailing the numerical procedure, we note that DoNOF also enables the computation of electrostatic moments through the keyword \texttt{IEMOM} in the \texttt{INPRUN} namelist, where \texttt{IEMOM=1} evaluates dipole moments (default), \texttt{IEMOM=2} additionally computes quadrupole moments, and \texttt{IEMOM=3} includes octopole moments as well. Furthermore, an external electric field may be specified explicitly through the keyword \texttt{EVEC}, which defines the three Cartesian components of the applied field in atomic units; by default all components are set to zero.

The NLOPs are evaluated numerically by applying the scaled fields and constructing Romberg-Richardson extrapolation triangles from centered finite-difference derivatives of the dipole moment. This approach enables the calculation of polarizabilities and hyperpolarizabilities directly within the NOF framework while avoiding the complexities of analytic response theory.

For a field applied along the $z$-axis, the response functions are defined as
\begin{equation}
\alpha_{zz} = \left.\frac{d\mu_z}{dF}\right|_{F=0}, \;
\beta_{zzz} = \left.\frac{d^{2}\mu_z}{dF^{2}}\right|_{F=0}, \;
\gamma_{zzzz} = \left.\frac{d^{3}\mu_z}{dF^{3}}\right|_{F=0}
\end{equation}
The electric-field strengths follow a dyadic ladder,
\begin{equation}
h_k = \texttt{STEP} \times 2^{k-1}, \qquad k = 1, \ldots, \texttt{NPOINT}
\end{equation}
and the dipole moment is evaluated at $\{-2h,\,-h,\, 0,\,h,\,2h \}$. Centered finite-difference formulas of order $O(h^2)$ are then used:
\begin{eqnarray}
\nonumber
\alpha(h) = \frac{\mu(+h) - \mu(-h)}{2h} \;\;\;\;\;\;\;\;\;\;\;\;\;\;\;\;\;\;\;\;\;\; \\ 
 \beta(h) = \frac{\mu(+h) - 2\mu(0) + \mu(-h)}{h^{2}} \;\;\;\;\;\;\;\;\;\;\;\;\;\; \\
\gamma(h) = \frac{\mu(+2h) - 2\mu(+h) + 2\mu(-h) - \mu(-2h)}{2h^{3}} 
\nonumber
\end{eqnarray}
To accelerate convergence toward the zero-field limit, DoNOF employs Romberg-Richardson extrapolation with $p=2$.\cite{Burden2011} The first column of the Romberg triangle is defined as
\begin{equation}
R(k,1) = D(h_k),
\end{equation}
where $D(h_k)$ denotes the finite-difference estimate of the desired property evaluated at step size $h_k$, i.e., $D(h_k)=\alpha(h_k)$, $\beta(h_k)$, or $\gamma(h_k)$. Higher-order estimates  ($j = 2,\ldots,\texttt{NPOINT}$) are then generated recursively as
\begin{equation}
R(k,j) = R(k,j-1) + \frac{R(k,j-1) - R(k+1,j-1)}{4^{\,j-1} - 1}
\end{equation}
with $k$ running up to $\texttt{NPOINT}-j+1$. The optimal extrapolated value is determined by scanning all vertical pairs in each column of the Romberg triangle and selecting the pair that minimizes 
\begin{equation}
\Delta_{k,j} = \left| R(k+1,j) - R(k,j) \right|.
\end{equation}
The selected value is then taken as $R(k,j+1)$, with $\Delta_{k,j}$ reported as the uncertainty. Ties are resolved by choosing the candidate of smallest absolute magnitude. A representative input illustrating how to compute polarizabilities and hyperpolarizabilities of \ce{H2} is shown below:

\begin{verbatim}
&INPRUN RUNTYP=`ENERGY' NLOP=-1 ERITYP=`FULL' /
$DATA
Title: H2 Nonlinear Optical Properties
aug-cc-pVQZ
H  1.0  0.0000  0.0000  -0.3707
H  1.0  0.0000  0.0000   0.3707
$END
&NOFINP IPNOF=8 NTHRESHL=5 NTHRESHE=10 /
\end{verbatim}

In this example, \texttt{NLOP=-1} activates the simultaneous calculation of $\alpha$, $\beta$, and $\gamma$. By default, \texttt{NPOINT=9} and \texttt{STEP =0.0001} define the dyadic field sequence used to build the Romberg-Richardson triangles. In addition, the convergence thresholds controlled by the keywords \texttt{NTHRESHL} and \texttt{NTHRESHE}, which monitor the symmetry of the Lagrange-multiplier matrix and the energy convergence, respectively, have been tightened to ensure stable and reliable values for the hyperpolarizabilities.

When the keyword \texttt{ISOALPHA=1} is specified with \texttt{NLOP=1}, DoNOF computes the diagonal components of the polarizability tensor by cycling the electric field over the $x$, $y$, and $z$ directions, yielding $\alpha_{xx}$, $\alpha_{yy}$, and $\alpha_{zz}$. The isotropic polarizability is then calculated as
\begin{equation}
\bar{\alpha} 
= \frac{\alpha_{xx} + \alpha_{yy} + \alpha_{zz}}{3},
\end{equation}
and the anisotropy (Raman convention) is given by
\begin{equation}
\Delta\alpha
= \sqrt{\frac{\left[(\alpha_{xx}-\alpha_{yy})^{2}+ (\alpha_{yy}-\alpha_{zz})^{2}
+ (\alpha_{zz}-\alpha_{xx})^{2}\right]}{2}} 
\end{equation}

DoNOF outputs the complete Romberg extrapolation tables for $\alpha$, $\beta$, and $\gamma$, the selected optimal values together with their estimated uncertainties, and, in the presence of \texttt{ISOALPHA=1}, the diagonal components of the polarizability tensor together with $\bar{\alpha}$ and $\Delta\alpha$.

\begin{table}[htbp]
    \caption{Components of the polarizability and second hyperpolarizabilities of the \ce{H2} molecule. }
    \bigskip
    \centering
    \begin{tabular}{c|c|c}
     Property       & GNOF & Exact\cite{Bishop1991} \\
    \hline
    $\alpha_{xx}$   &  4.58  & 4.58  \\
    $\alpha_{zz}$   &  6.40  & 6.39  \\
    $\bar{\alpha}$  &  5.19  & 5.18  \\
    $\Delta\alpha$  &  1.82  & 1.81  \\    
    $\gamma_{zzzz}$   & 696.7  & 682.5  \\
    \end{tabular}
    \label{tab:polH2}
\end{table}

Table \ref{tab:polH2} compares the polarizabilities and the second hyperpolarizability of \ce{H2} obtained with GNOF using the aug-cc-pVQZ basis set\cite{Kendall} against the essentially exact electronic results of Bishop et al.\cite{Bishop1991} derived from highly accurate explicitly correlated calculations. As expected for a centrosymmetric molecule, the first hyperpolarizability $\beta$ is identically zero and therefore not reported. The agreement between GNOF and the reference data is excellent for all listed properties. Both the transverse and longitudinal components of the static polarizability ($\alpha_{xx}=\alpha_{yy}$ and $\alpha_{zz}$) match the exact values to within 0.01 a.u., yielding correspondingly accurate isotropic average ($\bar{\alpha}$) and anisotropy ($\Delta\alpha$). The second hyperpolarizability $\gamma_{zzzz}$ is also reproduced with very good accuracy: the GNOF value differs from the benchmark by only about 2 percent, confirming that the functional provides a reliable description of higher-order response properties.

\section{Gradients, Hessian and Optimization}

Analytical gradients\cite{Mitxelena2017, Mitxelena2020c} with respect to the nuclear coordinates can be computed by setting \texttt{RUNTYP=`GRAD'}. Likewise, a numerical Hessian can be evaluated using \texttt{RUNTYP= `HESS'}. These capabilities allow for full geometry optimizations by selecting \texttt{RUNTYP=`OPTGEO'}, as illustrated in the following input:
\begin{verbatim}
 &INPRUN RUNTYP=`OPTGEO' MULT=1 ICHARG=0 /
 $DATA
 Title: Geometry Optimization of HCN
 cc-pVDZ
 H  1.0  0.000  0.000   1.064
 C  6.0  0.000  0.000   0.000
 N  7.0  0.000  0.000  -1.156
 $END
 &NOFINP IPNOF=8 /
\end{verbatim}

The results obtained for the geometry optimization of \ce{HCN} using the cc-pVDZ basis set are summarized in Table~\ref{tab:optgeo-hcn}. The \ce{HCN} molecule is linear, so its geometry is fully characterized by the \ce{H-C} and \ce{C-N} bond distances. Notably, the progression in correlation energy recovered by the different NOFs is evident in this example as well, with GNOFm yielding an energy at its optimal geometry that is comparable to the CCSD result.

\begin{table}[htbp]
    \caption{Comparison between optimized geometries and energies of \ce{HCN} computed with the cc-pVDZ basis set and the corresponding experimental values. \cite{Johnson2002-bo,Herzberg1950-bx}}   
    \includegraphics[scale=0.15]{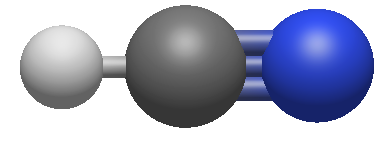}
    
    \vspace*{0.2cm}   
    \centering
    \begin{tabular}{l|ccc}
        \toprule
        \hline
        NOF     & H-C (\AA)   & C-N (\AA)   & Energy ($E_h$) \\
        \hline
        HF      & 1.067 & 1.134  & -92.884 \\
        PNOF5   & 1.064 & 1.154  & -92.987 \\
        PNOF7   & 1.055 & 1.170  & -93.032 \\
        GNOF    & 1.078 & 1.147  & -93.169 \\
        GNOFm   & 1.070 & 1.150  & -93.181 \\
        CCSD    & 1.080 & 1.168  & -93.181 \\
        \hline
        EXP     & 1.064 & 1.156  &    -     \\
        \hline
        \bottomrule
    \end{tabular}
    \label{tab:optgeo-hcn}
\end{table}

\section{NOF-based molecular dynamics}

With nuclear gradients available, ab initio molecular dynamics (AIMD) becomes feasible.\cite{Rivero-Santamaria2024-jd} Moreover, because a NOF incorporates static correlation, it can reliably describe non-equilibrium molecular structures and processes such as bond formation and breaking. Importantly, it enables real-time tracking of all orbitals throughout the simulation, providing valuable information on the ongoing nuclear dynamics, the evolution of chemical bonds, and key aspects of the reaction mechanism. \cite{Piris2024b} This contrasts with traditional multiconfigurational methods, which are typically restricted to limited active spaces whose relevant orbitals may change along the trajectory. AIMD simulations can be activated using the keyword \texttt{RUNTYP=`DYN’}, as illustrated in the following input:

\begin{verbatim}
 &INPRUN RUNTYP=`DYN' MULT=1 ICHARG=-1 /
 $DATA
 Title: Dynamics F- + H2 -> HF + H-
 cc-pVDZ
 F 9.0  -0.3000  -6.0000   0.0000
 H 1.0  -0.3707   0.0000   0.0000
 H 1.0   0.3707   0.0000   0.0000
 $END
 &NOFINP IPNOF=8 Imod=1 /
 &INPDYN dt=0.1 Vxyz=0,0.1,0,0.025,0,0,-0.025/
\end{verbatim}

This example corresponds to the \ce{F-} + \ce{H2} $\rightarrow$ HF + \ce{H-} reaction computed at the GNOFm/cc-pVDZ level of theory, which proceeds through an abstraction mechanism. The initial separation between the fluoride anion and the center-of-mass of \ce{H2} was set to 6 \AA\, and the anion was displaced 0.3 \AA\ to the left along the x-axis to introduce an impact parameter that facilitates the reaction. A time step of 0.1 fs was employed, ensuring reasonable conservation of the total energy.

In DoNOF, it is also possible to include an additional \texttt{INPDYN} namelist to define the conditions of the molecular dynamics, as shown in the example. This section specifies, for instance, the time step \texttt{dt} (in fs) and whether snapshots should be written at each step for orbital visualization by setting \texttt{snapshot=T}. Initial velocities in DoNOF can be defined in two ways. The first is by assigning values to the real two-dimensional array \texttt{Vxyz(1:3,1:NATOMS)}, which is set to zero by default. Alternatively, a trajectory file (.xyz) containing the initial positions and velocities of each atom may be provided by enabling the keyword \texttt{resflag=T}.

In our example, for simplicity, all velocities were set to zero except for that of the attacking anion, which was assigned a velocity of 0.1 \AA/fs along the $y$-axis, while the hydrogen atoms were given velocities of $\pm 0.025$ \AA/fs along the $x$-axis, corresponding to the zero-point vibrational motion of \ce{H2}. It is important to note that DoNOF operates in Cartesian coordinates; therefore, the initial kinetic energy is defined in the laboratory frame for \ce{F-}, which maps onto a specific collision energy in the center-of-mass frame. In this example, the resulting collision energy is $E_{\mathrm{col}} \approx 0.94\ \mathrm{eV}$.

\begin{figure}[htbp]
    \centering
    \includegraphics[width=0.48\textwidth]{H2F-dynamics.png}
    \caption{Selected snapshots from the \ce{F-} + \ce{H2} -> \ce{FH} + \ce{H-} dynamics. The bottom row indicates the time stamps corresponding to each snapshot.}
    \label{fig:nof-aimd-h2-f}
\end{figure}

Fig.~\ref{fig:nof-aimd-h2-f} shows the adiabatic evolution of a selected chemically active valence NO sampled every \texttt{dt=10fs} from \texttt{t=0fs} to \texttt{t=90fs}, along this reactive trajectory. This orbital describes both the breaking of the \ce{H2} $\Sigma$ bond and the formation of the F-H bond. Initially, the orbital corresponds to the sigma ``ss'' bonding orbital of singlet \ce{H2} combined with a 2p atomic orbital of the fluoride anion (\ce{F-}). As the reaction proceeds, it evolves into the $\sigma$ ``sp'' bonding orbital of singlet hydrogen fluoride (\ce{FH}) together with the $1s^2$ configuration of the hydride anion. Overall, this example illustrates how NOF-based AIMD provides insight into the evolution of electronic structure during chemical transformations, offering a valuable tool for elucidating complex reaction mechanisms.

\section{P\lowercase{y}NOF + D\lowercase{o}NOF.\lowercase{jl}}

While the Fortran version of DoNOF remains the primary and most efficient implementation, it is worth highlighting our experience translating the formalism into alternative programming languages. In this context, the PyNOF program was developed as a Python-based implementation of DoNOF, making use of the powerful tools provided by the Psi4NumPy project,\cite{Smith2018-ze} which internally relies on \emph{libint}.\cite{Libint2} A minimal input for performing a GNOF calculation with PyNOF is shown below:
\begin{verbatim}
  import pynof
  mol = pynof.molecule("""
  0 1
   O  0.0000   0.000   0.121
   H  0.0000   0.751  -0.485
   H  0.0000  -0.751  -0.485
   """)
   p = pynof.param(mol,"cc-pvdz")
   p.ipnof=8
   E,C,n = pynof.compute_energy(mol,p)
\end{verbatim}
The input code first specifies the molecular geometry, then constructs a parameter object, and finally invokes the energy evaluation. Under the hood, the implementation makes use of libraries such as \emph{numba}\cite{10.1145/2833157.2833162} for Just-In-Time compilation, \emph{numpy}\cite{harris2020array} for linear algebra and Einstein-summation operations, and \emph{cupy}\cite{Okuta2017CuPyA} to offload computationally intensive tasks, such as the orbital transformation, to GPUs. The Python version is intended as a platform for rapid prototyping of new equations, which can later be transferred into the more efficient, production-level Fortran code. An additional advantage is that it can be seamlessly integrated into Python-based workflows, including those commonly employed in machine learning and quantum computing applications.

Furthermore, we implemented the algorithm in the Julia language,\cite{bezanson2017julia} resulting in the DoNOF.jl package. Julia is a relatively recent programming language designed specifically for scientific computing, combining the ease of use of Python with performance comparable to compiled languages such as C or Fortran, thereby addressing the well-known two-language problem. The corresponding input for performing a GNOF single-point calculation in DoNOF.jl is shown below:
\begin{verbatim}
   using DoNOF
   mol = """
   0 1
    O  0.0000   0.000   0.121
    H  0.0000   0.751  -0.485
    H  0.0000  -0.751  -0.485
   """
   bset,p = DoNOF.molecule(mol,"cc-pvdz")
   p.ipnof = 8
   DoNOF.energy(bset,p)
\end{verbatim}

A close structural similarity with the PyNOF input can be observed. Einstein summation is handled through Tullio.jl and TensorOperations.jl,\cite{TensorOperations.jl} while GPU acceleration is enabled via CUDA.jl\cite{besard2018juliagpu,besard2019prototyping} and cuTENSOR.jl. One- and two-electron integrals are computed using the GaussianBasis.jl package, which internally relies on \emph{libcint}. As a result, the output files generated by DoNOF (Fortran) and DoNOF.jl (Julia) remain mutually compatible.

\section{Conclusions}

In this work, we present an updated release of the Donostia Natural Orbital Functional (DoNOF) program, together with new Python and Julia extensions. Combined with recent advances in natural-orbital-based functionals, the algorithms implemented in DoNOF firmly establish NOF-based computations as a robust alternative for addressing problems in physics and chemistry, including challenging cases involving strong correlation. We also expect this platform to serve as a valuable tool for functional developers, providing a practical environment in which new ideas can be designed, tested, and refined. Collaboration on the associated codebase is warmly encouraged.

\section{Acknowledgements}
Financial support comes from the Eusko Jaurlaritza (Basque Government), Ref.: IT1584-22, and from the Grant No. PID 2021-126714NB-I00, funded by MCIN/AEI/10.13039/501100011033. J. F. H. Lew-Yee acknowledges the DIPC and the MCIN program ``Severo Ochoa'' under reference AEI/CEX2018-000867-S for post-doctoral funding (Ref.: 2023/74.) The authors acknowledge the technical and human support provided by both the DIPC Supercomputing Center and IZO-SGI (SGIker) of EHU and European funding (ERDF and ESF).

\section{References}


%

\end{document}